\documentclass[conference]{IEEEtran}
\usepackage{cite}
\usepackage{amsmath,amssymb,amsfonts}
\usepackage{graphicx}
\usepackage{multirow}
\usepackage{subcaption}
\usepackage{url}
\usepackage[linesnumbered,ruled]{algorithm2e}
\def\BibTeX{{\rm B\kern-.05em{\sc i\kern-.025em b}\kern-.08em
    T\kern-.1667em\lower.7ex\hbox{E}\kern-.125emX}}

\begin{document}

\title{Parallel Longest Common SubSequence Analysis In Chapel}

\author{
\IEEEauthorblockN{Soroush Vahidi$^*$, Baruch Schieber$^*$, Zhihui Du$^+$,  David A. Bader$^+$}
\IEEEauthorblockA{\textit{$^*$Department of Computer Science} \\
\textit{$^+$Department of Data Science} \\
\textit{New Jersey Institute of Technology}\\
Newark, NJ, USA \\
\texttt{\{sv96,sbar,zd4,bader\}@njit.edu}}
}

\maketitle

\begin{abstract}
One of the most critical problems in the field of string algorithms is the longest common subsequence problem (LCS). The problem is NP-hard for an arbitrary number of strings but can be solved in polynomial time for a fixed number of strings. In this paper, we select a typical parallel LCS algorithm and integrate it into our large-scale string analysis algorithm library to support different types of large string analysis. Specifically, we take advantage of the high-level parallel language, Chapel, to integrate Lu and Liu's parallel LCS algorithm into Arkouda, an open-source framework. Through Arkouda, data scientists can easily handle large string analytics on the back-end high-performance computing resources from the front-end Python interface. The Chapel-enabled parallel LCS algorithm can identify the longest common subsequences of two strings, and experimental results are given to show how the number of parallel resources and the length of input strings can affect the algorithm's performance. 
\end{abstract}


\section{Introduction}
The longest Common Subsequence (LCS) of a set of strings is the longest string which is a subsequence of all of them. For example, the LCS of strings \emph{abccb}, \emph{abba} and \emph{acbb} is \emph{abb}. The finding of the LCS of some strings has applications, particularly in the context of bioinformatics, where strings represent DNA or protein sequences \cite{DJUKANOVIC2020106499}. 

Using a simple dynamic programming approach, one can find the LCS of two strings with lengths $m$ and $n$ in $\mathcal{O}(mn)$ time on one processor. For long strings, computing the LCS can take a long time, and researchers have tried to find faster algorithms. One way to increase the speed of the algorithm is to use an approximation algorithm instead of an exact algorithm. In this way, some methods, such as \cite{Ru19},\cite{akmal2021improved} and \cite{Be98}, have introduced approximation algorithms for the LCS problem and some of its variations. 

Another way to solve the LCS problem with a higher speed is to develop parallel algorithms. In Lu and Lin's work \cite{MI94}, two algorithms are suggested for finding the LCS of two strings in parallel, such that one of them has a time complexity of $\mathcal{O}(\log^2(m)+\log(n))$ with $mn/\log(m)$ processors, and the other one has time complexity $\mathcal{O}(\log^2(m)\log \log(m))$ with $mn/(\log^2(m)\log \log(m))$ processors. 

In this work, we have implemented a variant of the first algorithm and have measured its average running time for different test cases. To the best of our knowledge, it is the first parallel implementation of LCS in Chapel.
The main contributions in this paper are as follows:
\begin{enumerate}
    \item A typical longest common subsequence algorithm is implemented in Chapel to support high-performance string analysis. 
    \item Experimental results are given to show how the performance of the parallel algorithm will change with the size of two strings and the number of parallel resources. 
    \item This work is based on an open-source framework Arkouda \cite{merrill2019arkouda}. It means that data scientists can take advantage of the user-friendly Python language supported by Arkouda to conduct large-scale string analysis efficiently on the back-end high-performance computing resource with terabyte data or beyond. 
\end{enumerate}
 
\section{Algorithm Description and Parallel Implementation}

\subsection{Basic Idea}

Lu and Liu's parallel method, as presented in their work \cite{MI94}, offers a novel approach to solving the Longest Common Subsequence (LCS) problem. The central idea behind their method is to transform the LCS problem into a search for the maximum weighted path between two specially designated vertices within a grid graph.

In essence, this algorithm operates recursively. To elucidate, when tasked with discovering the maximum weighted path between vertices $a$ and $b$, it seeks out a strategic intermediary vertex, denoted as $c$. The objective is to maximize the combined weight of the path from $a$ to $c$ and the path from $c$ to $b$. Achieving this necessitates the determination of two critical components: the maximum weighted path from $a$ to $c$ and from $c$ to $b$. This recursive nature stems from the need to address these intermediate paths.

Like numerous other recursive algorithms, there exists a risk of exponential time complexity. To mitigate this concern, Lu and Liu's method employs dynamic programming techniques to efficiently tackle the recursive challenges posed by the problem. This pragmatic approach helps maintain computational tractability while deriving optimal solutions for the LCS problem.

\subsection{Recursive and Parallel Methods}

\subsubsection{Recursive Formula}
If vertices $a$ and $b$ are in two consecutive rows of the grid graph, the maximum weighted path between them can be calculated using a parallel prefix min algorithm.

In our implementation, we define several matrices, but for most of them, we compute only specific cells when needed. The most crucial matrix is denoted as $D_G$. Cell $(i,j)$ of $D_G$ indicates the column of the leftmost vertex in row $i$ of the grid graph $G$ that can be reached by a path with weight $j$ from the vertex in the $i$-th column of the first row. If matrix $D_G$ has more than two rows, we employ the following formula:
\[
D_G(i,j) = \min(D_{G_U}(i,j), D_{G_L}(i,j), D_{G_L}(D_{G_U}(i,k), j-k))
\]
for $1 \leq k \leq j$, where $D_{G_U}$ represents the upper half of $D_G$, and $D_{G_L}$ represents the lower half of $D_G$.

To better understand this formula, consider that $D_{G_U}(i,k)$ represents the leftmost vertex in the bottom row of $G_U$ that can be reached from the $i$-th vertex of the first row of $G_U$ with a path of weight $k$. Therefore, $D_{G_L}(D_{G_U}(i,k))(j-k)$ represents the leftmost vertex in the bottom row of $G_U$ that can be reached from the $i$-th vertex of the first row of $G$ such that the weight of this path is $j$. The sum of the weight of the edges in $G_U$ for this path is $k$, and the sum of the weights of the edges in $G_L$ for this path is $j-k$. An example of computing a cell of $D_G$ is shown in Figure \ref{fig:dgu}.
\begin{figure}[htp]
    \centering
    \includegraphics[width=8cm]{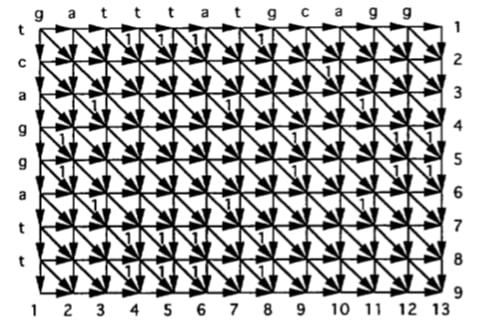}
    \caption{Finding the LCS of \emph{gatttatgcagg} and \emph{tcaggatt} is equal to finding the maximum weighted path in this graph from the upper left vertex to downright vertex. This figure is copied from \cite{MI94}. It's worth noting that in our implementation, we exclude diagonal edges with a weight of 0.}
    \label{fig:example1}
\end{figure} 

We define a vertex $v$ in the bottom row of a grid graph $G$ as the $j$-th breakout vertex of the vertex $G(1, i)$ if $v$ is the leftmost vertex in the bottom row and there exists a path with cost $j$ from vertex $G(1, i)$ to $v$. For instance, in Figure \ref{fig:example1}, vertices (9, 2), (9, 3), (9, 4), (9, 5), and (9, 13) represent the first, second, third, fourth, and fifth breakout vertices of the vertex (1,1). There is no $5^{th}$ breakout for vertex (1,8), or we can say that the 5th breakout of vertex (1,8) is $\infty$. An example of computing a cell of $D_G$ can be seen in Figure \ref{fig:dgu}.

A matrix is considered monotone if, given two consecutive columns $c_1$ and $c_2$ with $c_1$ to the left of $c_2$, the cell with the minimum value in $c_2$ is not in a row higher than the row containing the cell with the minimum value in $c_1$.

\begin{figure}[htp]
    \centering
    \includegraphics[width=8cm]{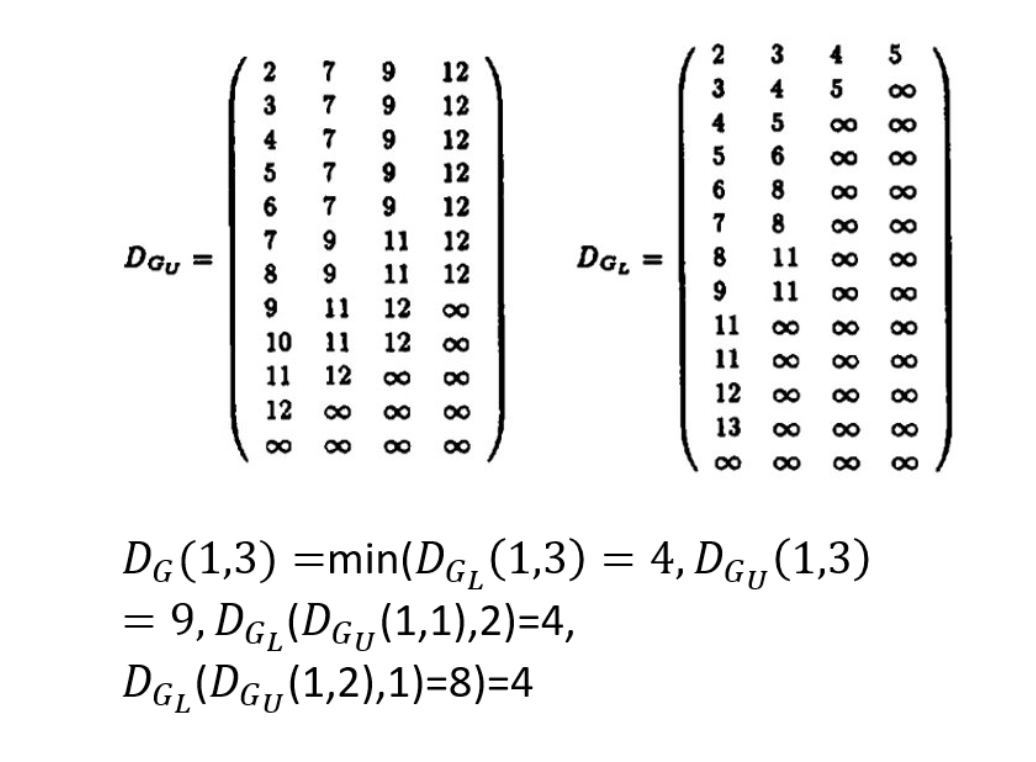}
    \caption{$D_{G_U},D_{G_L}$ and computing $D_G(1,3)$ for the graph in Fig \ref{fig:example1}. This figure is copied from \cite{MI94}.}
    \label{fig:dgu}
\end{figure}

\subsubsection{Key Functions}

\begin{algorithm}[tpbh]
\small
\DontPrintSemicolon
\LinesNumbered
\caption{Find ColMins Function}
\label{alg:colmin}
$findColMins(dgu,dgl,vertex:int, left: int, right: int, top: int, bottom: int, mins,firstind:int) $ \\
    $var$ $ cols = right - left + 1$\; 
    \If {($cols < 1$)}{ 
      \Return \;
    }
    $var$ $midCol = \lceil(right+left)/2\rceil : int$\;
    $var$  $minIndex = findMinIndex(dgu,dgl,vertex:int, midCol, top, bottom)$ \;
    $mins[firstind+midCol-left] = minIndex$\;
    \If { ($find\_cell(dgu,dgl,vertex,minIndex,midCol)\ne infin$)} {  
         $cobegin$\{ \;
            \hspace{8pt}   $findColMins(dgu,dgl,vertex:int, left, midCol-1, top, minIndex, mins,firstind)$ \; 
             \hspace{8pt} $findColMins(dgu,dgl,vertex:int, midCol+1, right, minIndex, bottom, mins,firstind+midCol-left+1)$ \;  
        \}\;
    \Else  { 
      $findColMins(dgu,dgl,vertex:int, left, midCol-1, top, bottom, mins,firstind)$;
      }
    }
\end{algorithm}

\begin{algorithm}[tpbh]
\small
\DontPrintSemicolon
\LinesNumbered
\caption{Find Min Index Function}
\label{alg:minind}
$findMinIndex( dgu,dgl,vertex:int, col: int, top: int, bottom: int) $ \\
    $var$ $  listsize = bottom - top + 1$ \;
    $var$ $exp: int = 1$\;
    $var$ $expm1, expnot: int$\;
    $var$ $prefix: [0..listsize-1] int$\;
    $var$ $minIndex: [0..listsize-1] int$\;
    \ForAll {(i in 0..listsize-1)} {
      $prefix[i] = find\_cell(dgu,dgl,vertex,i+top,col)$\; 
      $minIndex[i] = i$\;
    }
    
    \While {(exp $<$ listsize)} {
      $expm1 = exp - 1$\;
      $expnot = \tilde exp$\;
      \ForAll {(j in 0..listsize-1)} {
        \If {($j\&exp \neq 0$)}{
          \If {($prefix[j\&expnot|expm1] \leq prefix[j]$ )} {
            $prefix[j] = prefix[j\&expnot|expm1]$ \;
            $minIndex[j] = minIndex[j\&expnot\|expm1]$ \;
          }
        }
      }
      $exp = exp < < 1$\;
  }
  \Return $minIndex[listsize-1]+top$ \;
\end{algorithm}
A critical component of the matrix computation for $D_G$ (the cost matrix of $G$) lies in Algorithm \ref{alg:colmin}. This algorithm recursively identifies the minimum element index in each monotone matrix column and stores these indices in an array called $mins$. Specifically, $mins[i]$ preserves the index of the minimum element in column $i$. The variables $left$, $right$, $top$, and $bottom$ correspond to the first and last columns and the matrix's first and last rows, respectively.

We consistently initialize the variable $firstind$ to match the value of $left$. Cases where $left \neq firstind$ arise in recursive processes, but these intricacies do not require user intervention. Within the pseudocode, the commands $forall$ and $cobegin$ signify situations where all enclosed commands will be executed in parallel, while $for$ executes commands within its loop sequentially.

Algorithm \ref{alg:colmin} presents the pseudocode for $ColMin$. Inside Algorithm \ref{alg:colmin}, we rely on Algorithm \ref{alg:minind}, which is responsible for determining the index of the minimum value within a column of a matrix. Notably, Algorithm \ref{alg:minind} operates in parallel.

It's important to note that every recursive relation possesses its own set of initial values.

\subsubsection{Computing $D_{G_H}$}
In the recursive computation of $D_G$, we do not recursively compute the cost matrix of $G$ if $G$ has only two rows; instead, we approach it differently. We consider the input strings $a$ and $b$ and define the cost matrix of the grid graph $G$ consisting of rows numbered $h$ and $h+1$ as $D_{G_h}$.

In our implementation, we assume that $D_{G_H}$ has only two rows. The first row of this matrix represents the $0^{th}$ breakout for each vertex, and we define the $0^{th}$ breakout of the $i^{th}$ vertex of $G_h$ as $i$. While \cite{MI94} does not define the $0^{th}$ breakout, we introduce this definition for simplifying the implementation. Therefore, for the sake of simplicity in notation, we assume that $D_{G_h}$ consists of only one row, representing the first breakout of each vertex in the upper row of the grid graph comprising rows $h$ and $h+1$ of $G$.

We represent the $i^{th}$ letter of the string $s$ as $s_i$, where $i \geq 1$. To compute $D_{G_H}$, we need to determine values $j_1, j_2, j_3,\ldots, j_r$ such that $j_1 < j_2 < j_3 < \ldots < j_r$, and $b_{j_i} = a_h$ for $1 \leq i \leq r$. Finding these values can be accomplished in $\mathcal{O}(\log n)$ using $n$ processors, where $n$ represents the size of string $b$. Afterward, we assign $j_{k} - j_{k-1}$ to $D_{G_H}(j_{k-1}+1)$ for $1 < k \leq r$, and set $D_{G_h}(1)$ to $j_1+1$. 

For instance, after performing these steps to compute $D_{G_1}$ in Fig. \ref{fig:example1}, we obtain $j = (3,4,5,7)$ and $D_{(G_1)} = (4, x, x, 1, 1, 2, x, x, x, x, x, x)$, where $x$ represents values that have not yet been computed. Subsequently, we set $D_{G_h}(k) = \sum_{j=1}^k D_{G_h}(j)$ for $1 \leq k \leq j_r+1$. At the conclusion of this step, $D_{G_1} = (4,4,4,5,6,8,8,8,8,8,8,8)$. In the final step of computing $D_{G_h}$, we assign $\infty$ to the entries $j_r+2$ to $n$ of $D_{G_h}$. Consequently, we arrive at $D_{G_1} = (4,4,4,5,6,8,8,\infty,\infty,\infty,\infty,\infty)$.

The computation of $D_{G_h}$ for all values of $h$ can be achieved in $\mathcal{O}(\log n)$ using $mn/\log(n)$ processors \cite{MI94}.

\subsection{Finding the Maximum Weighted Path}
After computing matrix $D_G$, which represents the weights of various paths, we need to extract the vertices of the maximum weighted path from the upper-left vertex (referred to as the source) of $G$ to the lower-right vertex (referred to as the sink). For a maximum-cost path $P=\langle v_1,v_2,\ldots,v_l \rangle$ from the source to the sink in $G$, there can be multiple vertices in $P$ that belong to the same row in $G$.

A vertex $v_i$ in $P$ is considered a cross-vertex if it is the leftmost vertex of $P$ within its respective row. We use the notation $v[j]$ to represent a cross-vertex on the $j$-th row of $G$, distinguishing it from other vertices in $P$. It is evident that $v_1 = v[1]$, assuming row number 1 (not 0) as the first row.

\subsection{Eliminating LCS from $D_G$}
Now, we need to address two subproblems: identifying the cross-vertices of $P$ and identifying the other vertices of $P$. Let's start with the first subproblem:

All cross-vertices on a maximum-cost path can be determined as a byproduct of computing the cost matrix $D_G$. Suppose we are computing $D_G(i,j)$, which corresponds to finding in $G$ the $j$-th breakout vertex of $x = G(1,i)$, denoted as $y$. Let $p$ be the maximum-cost path from $x$ to $y$, and let vertex $q$ be the cross-vertex of $p$ on the boundary between $G_U$ and $G_L$. This implies that $q = v[m/2+1]$.

The second subproblem is straightforward. If $v[i]$ and $v[i+1]$ represent vertex $G(i,j_1)$ and vertex $G(i+1,j_2)$, respectively, then the vertices on the $i$-th row of $G$ from $G(i,j_1+1)$ to $G(i,j_2-1)$ must all be part of the vertices between $v[i]$ and $v[i+1]$ in $p$, considering that diagonal edges with weight 0 are not considered. Therefore, once all cross-vertices have been identified in the first stage, there should be no difficulty in listing all the vertices of $p$ in an array. This can be accomplished using a parallel PrefixSum function with a time complexity of $\mathcal{O}(\log n)$, employing $n$ processors.

\subsection{Identifying the LCS}
In the final stage of the algorithm, we examine the cost of each edge $e=(v[k],v[k+1])$. Symbol $a_i$ is marked if we find that the edge $e$ has a cost of 1 and vertex $v[k]$ has a column index of $i$.
The LCS of strings $a$ and $b$ corresponding to path $p$ can be obtained by sorting these marked symbols. Given that the number of edges on $p$ is bounded by $n + m$, and checking the cost of an edge takes constant time, marking symbols in $a$ can be accomplished in constant time using $n$ processors or in $\mathcal{O}(\log n)$ using $n/(\log n)$ processors.


\section{Experimental Results}
\subsection{Experimental System}

We conducted our experiments on a system with 2.00GHz Intel(R) Xeon(R) Gold 6330 CPUs. Our program was executed using Chapel version 1.31.0.

In our Chapel configuration, we set the \texttt{CHPL\_TASKS} variable to \texttt{``qthreads''}, and \texttt{CHPL\_LLVM} was configured as \texttt{``bundled''}. The number of cores we utilized was controlled using the command \texttt{export CHPL\_RT\_NUM\_THREADS\_PER\_LOCALE=x}, where $x$ represents the desired number of cores.


\subsection{Performance}
\begin{figure}[htp]
    \centering
    \includegraphics[width=8cm]{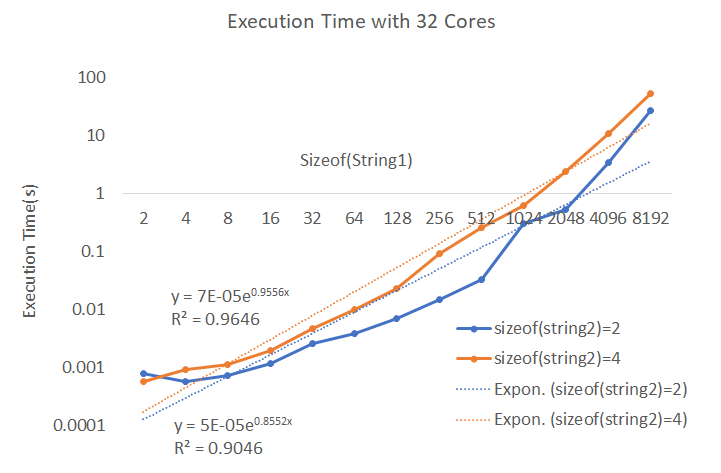}
    \caption{The execution time experiences exponential growth as one string's size increases while the other remains fixed. This phenomenon occurs within the context of a computational environment equipped with a total of 32 cores.}
    \label{fig:32core}
\end{figure}

In this section, we embark on an in-depth exploration of the multifaceted performance characteristics exhibited by the proposed parallel algorithm. Our initial focus is on examining how the execution time is influenced by varying the lengths of input strings, with one of them held constant. The comprehensive results of these investigations are meticulously presented in Fig. \ref{fig:32core}.

Fig. \ref{fig:32core} eloquently illustrates a series of experiments where we meticulously maintain the length of one string at values of 2 and 4, while systematically extending the size of the other string from 2 to 8192. These empirical investigations were conducted with 32 processing cores.

Our observations from this figure reveal a striking pattern of nearly exponential growth in the total execution time required to determine the longest common subsequence. This growth is prominently evident when we hold the length of one string constant and progressively vary the length of the other. Specifically, when one string size is fixed at 2, our rigorous analysis yields a precise regression equation of $ time = 5 \times 10^{-05} \times e^{0.8552 \times \text{size}}$, accompanied by an $R^2$ value of 0.9046. Similarly, for the scenario where one string length remains constant at 4, our analysis furnishes the regression equation as $ time = 7 \times 10^{-05} \times e^{0.9556 \times \text{size}}$, accompanied by a notably higher $R^2$ value of 0.9646.

These findings distinctly underscore the algorithm's remarkable sensitivity to input size. This sensitivity is vividly exemplified by the substantial and expedited growth in execution time experienced when dealing with larger strings.

\begin{figure}[htp]
    \centering
    \includegraphics[width=8cm]{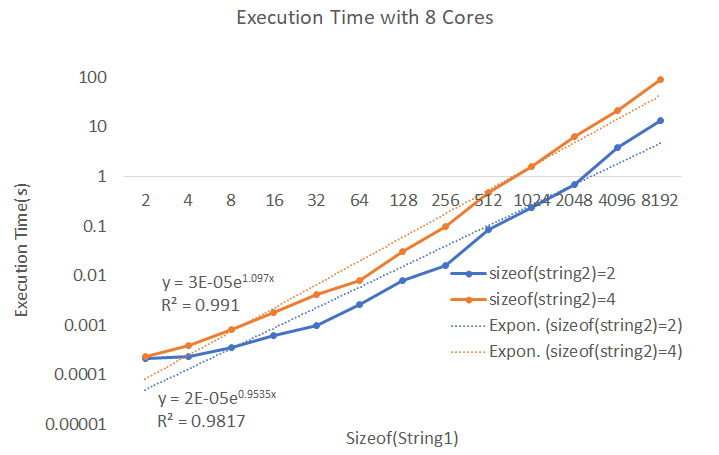}
    \caption{The execution time experiences exponential growth as one string's size increases while the other remains fixed. This phenomenon occurs when we reduce the number of cores from 32 to 8.}
    \label{fig:8core}
\end{figure}

Intriguingly, as we look at the results obtained with eight processing cores (as depicted in Fig. \ref{fig:8core}), we discern a similar trend. However, subtle differences emerge when examining the fitting equations. When one string length is kept at 2, our analysis yields a fitting equation of $ time = 2 \times 10^{-05} \times e^{0.9535 \times \text{size}}$, resulting in an exceptionally high $R^2$ value of 0.9817. Similarly, for a fixed string length of 4, the regression equation is expressed as $ time = 3 \times 10^{-05} \times e^{1.097 \times \text{size}}$, with an even higher $R^2$ value of 0.991.

These nuances in the results with eight cores highlight that (1) for the same fixed string size, increasing the size of the other string incurs a significantly faster growth in execution time. Notably, focusing on the exponent constants reveals that, for a fixed string size of 2, the execution time increase with 8 cores is approximately $0.4 \times e^{0.1283}$ times that of 32 cores. Similarly, for a fixed string size of 4, the execution time increase with 8 cores is roughly $\frac{3}{7} \times e^{0.1414}$ times that of 32 cores. These insights underscore the intriguing relationship between input length and core count, elucidating that increasing string length has a more profound impact on execution time than reducing the number of processing cores.

In Table \ref{tab:comparison}, we expand our testing to include various fixed sizes of strings. Then, we calculate the speedup of performance on 32 cores compared to that on 8 cores. The results underscore two key observations:

(1) Effective Parallelization: As we add processing cores, a clear reduction in total execution time becomes evident for identical string sizes. When both string lengths are larger, the evidence becomes more obvious. An average of $1.8\times$  speedup can be achieved. This demonstrates the tangible effectiveness of our parallel method, affirming its ability to optimize performance.

(2) Input Size Impact: It is noteworthy that amplifying the lengths of either string substantially impacts the total execution time. Specifically, an increase in the size of either string leads to a noticeable escalation in the overall execution duration.

These insights provide valuable confirmation of the efficacy of our parallel approach while highlighting the sensitivity of execution time to changes in input size.

\begin{table*}[]
\caption{Algorithm execution time (seconds) and speedup for different number of cores and string sizes}
\label{tab:comparison}
\begin{tabular}{|c|c|cccc|cccc|}
\hline
\multirow{2}{*}{number   of cores} & \multirow{2}{*}{sizeof(string1)} & \multicolumn{4}{c|}{sizeof(string2)}                                                                  & \multicolumn{4}{c|}{Speedup}                                                                             \\ \cline{3-10} 
                                   &                                  & \multicolumn{1}{c|}{2}        & \multicolumn{1}{c|}{4}       & \multicolumn{1}{c|}{8}       & 16      & \multicolumn{1}{c|}{2}        & \multicolumn{1}{c|}{4}        & \multicolumn{1}{c|}{8}        & 16       \\ \hline
\multirow{4}{*}{32}                & 1024                             & \multicolumn{1}{c|}{0.119569} & \multicolumn{1}{c|}{0.62156} & \multicolumn{1}{c|}{1.63747} & 3.2784  & \multicolumn{1}{c|}{3.095577} & \multicolumn{1}{c|}{2.555988} & \multicolumn{1}{c|}{1.406334} & 1.449948 \\ \cline{2-10} 
                                   & 2048                             & \multicolumn{1}{c|}{0.890801} & \multicolumn{1}{c|}{2.4456}  & \multicolumn{1}{c|}{4.51969} & 12.2226 & \multicolumn{1}{c|}{1.54443}  & \multicolumn{1}{c|}{2.583583} & \multicolumn{1}{c|}{1.777177} & 1.575164 \\ \cline{2-10} 
                                   & 4096                             & \multicolumn{1}{c|}{6.64456}  & \multicolumn{1}{c|}{11.1501} & \multicolumn{1}{c|}{16.5999} & 51.9423 & \multicolumn{1}{c|}{0.805853} & \multicolumn{1}{c|}{1.902557} & \multicolumn{1}{c|}{1.997831} & 1.40539  \\ \cline{2-10} 
                                   & 8192                             & \multicolumn{1}{c|}{29.5529}  & \multicolumn{1}{c|}{53.5266} & \multicolumn{1}{c|}{67.5972} & 190.575 & \multicolumn{1}{c|}{0.971884} & \multicolumn{1}{c|}{1.648791} & \multicolumn{1}{c|}{2.152708} & 2.1932   \\ \hline
\multirow{4}{*}{8}                 & 1024                             & \multicolumn{1}{c|}{0.370135} & \multicolumn{1}{c|}{1.5887}  & \multicolumn{1}{c|}{2.30283} & 4.75351 & \multicolumn{4}{c|}{\multirow{4}{*}{}}                                                                   \\ \cline{2-6}
                                   & 2048                             & \multicolumn{1}{c|}{1.37578}  & \multicolumn{1}{c|}{6.31841} & \multicolumn{1}{c|}{8.03229} & 19.2526 & \multicolumn{4}{c|}{}                                                                                    \\ \cline{2-6}
                                   & 4096                             & \multicolumn{1}{c|}{5.35454}  & \multicolumn{1}{c|}{21.2137} & \multicolumn{1}{c|}{33.1638} & 72.9992 & \multicolumn{4}{c|}{}                                                                                    \\ \cline{2-6}
                                   & 8192                             & \multicolumn{1}{c|}{28.722}   & \multicolumn{1}{c|}{88.2542} & \multicolumn{1}{c|}{145.517} & 417.969 & \multicolumn{4}{c|}{}                                                                                    \\ \hline
\end{tabular}
\end{table*}

\section{Related Work}
\looseness=-1
In \cite{Yang2010AnEP}, Yang \textit{et al.} developed an efficient parallel algorithm on GPUs for the LCS problem. They proposed a new technique that changes the data dependency in the score table used by dynamic programming algorithms to enable higher degrees of parallelism.
In \cite{Garcia03}, Garcia~\emph{et al.} introduce a coarse-grained multicomputer algorithm which works in $\mathcal{O}(N^2/P)$ time complexity with $P$ processors and $\mathcal{O}(P)$ communication steps.
Dhraief \emph{et al.} \cite{Dhraief2011ParallelCT} studied languages for parallel development on GPUs (CUDA and OpenCL) and presented a parallelization approach to solving the LCS problem on GPU. Their proposed algorithm was evaluated on an NVIDIA platform using CUDA and OpenCL.
Babu \emph{et al.} \cite{klo08} introduced a parallel algorithm to compute the LCS using graphics hardware acceleration and multiple levels of parallelism.
Babu and Saxena \cite{Bab97} introduced an algorithm with $\mathcal{O}(\log m)$ time complexity using $mn$ processors, where $m$ is the length of the shorter string and $n$ is the length of the longer string.
Several parallel algorithms (e.g., \cite{K08}, \cite{Q11}, and \cite{Chen2006}) have been proposed to find the LCS of multiple strings.
Nguyen \emph{et al.} \cite{gpugem} introduced the basics of parallel prefix scans.
Tchendji~\emph{et al.} \cite{TCHENDJI20} provided a parallel algorithm to solve the LCS problem with constraints. Specifically, their problem is to find the longest common subsequence, which excludes some strings as its substrings.
In  \cite{Alves2006}, Alves, Caceres and Song introduce a parallel algorithm for the all-substrings longest common subsequence problem. In this problem, given two strings $A$ and $B$, the goal is to compute the LCS of $A$ and each substring of $B$ denoted as $B'$.

\section{Conclusions and Future work}

This paper introduces a parallel algorithm implementation for calculating the Longest Common Subsequence (LCS) of two strings using the Chapel programming language.  It includes an analysis of the algorithm's average runtime across strings of varying lengths on different numbers of cores.
Our source code is open source and available on GitHub at
\url{https://github.com/SoroushVahidi/parallel-longest-common-subsequence/}

Our future research endeavors will focus on expanding the capabilities of this algorithm implementation. Specifically, we plan to develop a comprehensive library for LCS computation in Chapel, which will encompass additional parallel methods for solving the LCS problem.

Furthermore, an intriguing avenue for future research lies in the development of parallel algorithms tailored to address various LCS problems with specific constraints. These efforts aim to provide more versatile and efficient solutions for a wide range of real-world applications.

 \section*{Acknowledgment}
We thank the Chapel and Arkouda communities for their support, as well as the NSF funding support through grant CCF-2109988. We also appreciate Nese L. Us for helping us debug the code and Jose L.  Mojica Perez for helping us debug the code and install Chapel.

\bibliographystyle{plain}
\bibliography{ref,arkouda-chapel}

\end{document}